\documentclass{elsart}

\usepackage{epsfig}

\newcommand{\be}{\begin{equation}}
\newcommand{\ee}{\end{equation}}
\newcommand{\bea}{\begin{eqnarray}}
\newcommand{\eea}{\end{eqnarray}}

 \def\bean{\begin{eqnarray*}}
 \def\eean{\end{eqnarray*}}

 \def\l{\left}
 \def\r{\right}
 \def\Im{{\rm Im}}
 \def\Re{{\rm Re}}
 \def\bm#1{\mbox{\boldmath$#1$}}
 \def\gsim{\mathrel{\rlap{\lower0.2em\hbox{$\sim$}}\raise0.2em\hbox{$>$}}}
 \def\ksim{\mathrel{\rlap{\lower0.2em\hbox{$\sim$}}\raise0.2em\hbox{$<$}}}

\begin{document}

\begin{frontmatter}
\title{QCD thermodynamics and confinement from a dynamical quasiparticle point of
view}
\author[unig]{W. Cassing\corauthref{cor1}}
\ead{Wolfgang.Cassing@theo.physik.uni-giessen.de}
\corauth[cor1]{corresponding author}
\address[unig]{Institut f\"ur Theoretische Physik, %
  Universit\"at Giessen, %
  Heinrich--Buff--Ring 16, %
  D--35392 Giessen, %
  Germany}

\begin{abstract}
  In this study it is demonstrated that a simple picture of the
  QCD gluon liquid emerges in the dynamical quasiparticle model
  that specifies the active degrees of
  freedom in the time-like sector and yields a potential energy
  density in the space-like sector.  By using the
  time-like gluon density (or scalar gluon density)
  as an independent degree of freedom -
  instead of the temperature $T$ as a Lagrange parameter -
  variations of the potential energy density lead to effective
  mean-fields for time-like gluons and an effective gluon-gluon interaction
  strength at low density. The latter yields a simple dynamical picture for
  the gluon fusion to color neutral glueballs when approaching the phase boundary
  from a temperature higher than $T_c$ and paves the way for an
  off-shell transport theoretical description of the parton
  dynamics.

\end{abstract}

\begin{keyword}
Quark gluon plasma, General properties of QCD, Relativistic
heavy-ion collisions \PACS 12.38.Mh\sep 12.38.Aw\sep 25.75.-q
\end{keyword}

\end{frontmatter}

%\newpage

\section{Introduction}
The formation of a quark-gluon plasma (QGP) and its transition to
interacting hadronic matter -- as occurred in the early universe
-- has motivated a large community for several decades (cf.\
\cite{QM01} and Refs.\ therein). Early concepts of the QGP were
guided by the idea of a weakly interacting system of partons
(quarks, antiquarks and gluons) since the entropy $s$ and energy
density $\epsilon$ were found in lattice QCD to be close to the
Stefan Boltzmann (SB) limit for a relativistic noninteracting
system \cite{Karsch}. However, this notion had to be given up in
the last years since experimental observations at the Relativistic
Heavy Ion Collider (RHIC) indicated that the new medium created in
ultrarelativistic Au+Au collisions was interacting more strongly
than hadronic matter. Moreover, in line with earlier theoretical
studies in Refs. \cite{Thoma,Andre,Shuryak} the medium showed
phenomena of an almost perfect liquid of partons \cite{STARS,Miklos3} as
extracted from the strong radial expansion and elliptic flow of
hadrons as well the scaling of the elliptic flow with parton
number {\it etc}. The latter collective observables have been
severely underestimated in conventional string/hadron transport
models \cite{Cassing03,Brat04,Cassing04}, but hydrodynamical
approaches did quite well in describing (at midrapidity) the
collective properties of the medium generated during the early
times for low and moderate transverse momenta \cite{Heinz,Bass2}.
Soon the question came up about the constituents of this liquid;
it might be some kind of i) "epoxy" \cite{GerryEd}, i.e. a system
of resonant or bound gluonic states with large scattering length,
ii) a system of chirally restored mesons, instanton molecules or
equivalently giant collective modes \cite{GerryRho}, iii) a system
of colored bound states of quarks $q$ and gluons $g$, i.e.\ $gq$,
$qq$, $gg$ etc.\ \cite{Eddi}, iv) some 'string spaghetti' or
'pasta' {\it etc}. In short,  many properties of the new phase are
still under debate and practically no dynamical concepts are
available to describe the freezeout of partons to color neutral
hadrons that are subject to experimental detection.

Lattice QCD (lQCD) calculations provide some guidance to the
thermodynamic properties of the partonic medium close to the
transition at a critical temperature $T_c$ up to a few times
$T_c$, but lQCD calculations for transport coefficients presently
are not accurate enough \cite{lattice2} to allow for firm
conclusions. Furthermore, it is not clear whether the partonic
system really reaches thermal and chemical equilibrium in
ultrarelativistic nucleus-nucleus collisions and nonequilibrium
models are needed to trace the entire collision history.  The
available string/hadron transport models
\cite{Cass99,URQMD1,URQMD2} are not accurate enough - as pointed
out above - nor do partonic cascade simulations
\cite{Geiger,Zhang,Molnar,Bass} (propagating massless partons)
sufficiently describe the reaction dynamics when employing cross
sections from perturbative QCD (pQCD). This also holds - to some
extent -  for the Multiphase Transport Model AMPT \cite{AMPT}
since it includes only on-shell massless partons in the partonic
phase as in Ref. \cite{Zhang}. The same problem comes about in the
parton cascade model of Xu and Greiner \cite{Carsten} where
additional 2$ \leftrightarrow$ 3 processes like $gg
\leftrightarrow ggg$ are incorporated. On the other hand it is
well known that strongly interacting quantum systems require
descriptions in terms of propagators $D$ with sizeable
selfenergies $\Pi$ for the relevant degrees of freedom. Whereas
the real part of the selfenergy gives contributions to the energy
density, the imaginary parts of $\Pi$ provide information about
the lifetime and/or reaction rate of time-like 'particles'
\cite{Andre}. In principle, off-shell transport equations are
available in the literature \cite{Juchem,Sascha1,Leo}, but have been applied
only to dynamical problems where the width of the quasiparticles
stays moderate with respect to the pole mass \cite{Laura}. On the
other hand, the studies of Peshier \cite{Andre04,Andre05} indicate
that the effective degrees of freedom in a partonic phase should
have a width $\gamma$  in the order of the pole mass $M$ already
slightly above $T_c$.

The present study addresses essentially three questions: i) Do we
understand the QCD thermodynamics in terms of dynamical
quasiparticles down to the phase boundary in a 'top down' scenario
and what are the effective degrees of freedom as well as energy
contributions? ii) Can such a quasiparticle approach help in
defining an off-shell transport model that - at least in thermal
equilibrium - reproduces the thermodynamic results from lQCD? iii)
Are there any perspectives in modeling the transition from
partonic to hadronic degrees of freedom in a dynamical way?

The present work is exploratory in the sense that it is restricted
to a pure gluonic system of $N_c^2-1$ gluons with two transverse
polarisations, i.e. degeneracy $d_g$ = 16 for the gluonic
quasiparticles that are treated as relativistic scalar fields.
Note, however, that the qualitative features stay the same when
adding light quark degrees of freedom \cite{Andre05}; this finding
is well in line with the approximate scaling of thermodynamic
quantities from lQCD when dividing by the number of degrees of
freedom and scaling by the individual critical temperature $T_c$
which is a function of the different number of parton species
\cite{Karsch5}.

The outline of the paper is as follows: After a short
recapitulation of the dynamical quasiparticle model in Section 2
new results on the space-like and time-like parts of observables
are presented that allow for a transparent physical
interpretation. In Section 3 we will examine derivatives of the
space-like part of the quasiparticle energy density with respect
to the time-like (or scalar) density which provides information on
gluonic mean fields and their effective interaction strength. The
implications of these findings with respect to an off-shell
transport description are pointed out throughout the study. A summary
and extended discussion closes this work in Section 4.

\section{Off-shell elements in the DQPM}
\subsection{Reminder of the DQPM}
The Dynamical QuasiParticle Model (DQPM)\footnote{DQPM also stands
alternatively for Dynamical-Quasiparticle-Peshier-Model} adopted
here goes back to Peshier \cite{Andre04,Andre05} and starts with
the entropy density $s$ in the quasiparticle limit ~\cite{BlaizIR},
\be
  s^{dqp}
  =
  - d_g\!\int\!\!\frac{d \omega}{2 \pi} \frac{d^3p}{(2 \pi)^3}
  \frac{\partial n}{\partial T}
   \l( \Im\ln(-\Delta^{-1}) + \Im\Pi\,\Re\Delta \r)\!,
  \label{sdqp}
\ee where $n(\omega/T) = (\exp(\omega/T)-1)^{-1}$ denotes the Bose
distribution function, $\Delta$ stands for the scalar
quasiparticle propagator and $\Pi$ for the quasiparticle
selfenergy which is considered here to be a Lorentz scalar. In
principle, the latter quantities are Lorentz tensors and should be
evaluated in a nonperturbative framework. However, a more
practical procedure is to use a physically motivated {\em Ansatz}
with a Lorentzian spectral function,
\be
 \rho(\omega)
 =
 \frac\gamma{ E} \l(
   \frac1{(\omega-E)^2+\gamma^2} - \frac1{(\omega+E)^2+\gamma^2}
 \r) ,
 \label{eq:rho}
\ee and to fit the few parameters to results from lQCD. With the
convention $E^2(\bm p) = \bm p^2+M^2-\gamma^2$, the parameters
$M^2$ and $\gamma$ are directly related to the real and imaginary
parts of the corresponding (retarded) self-energy, $\Pi =
M^2-2i\gamma\omega$. It should be stressed that the entropy
density functional (\ref{sdqp}) is not restricted to  quasiparticles
of low width $\gamma$ and thus weakly interacting particles. In
fact, in the following it will be shown that a novel picture of
the hot gluon liquid emerges because $\gamma$ becomes comparable
to the quasiparticle mass already slightly above $T_c$
\cite{Andre04,Andre05}.

\begin{figure}[htb!]
  \begin{center}
  \vspace{0.1cm}
    \includegraphics[width=11.5cm]{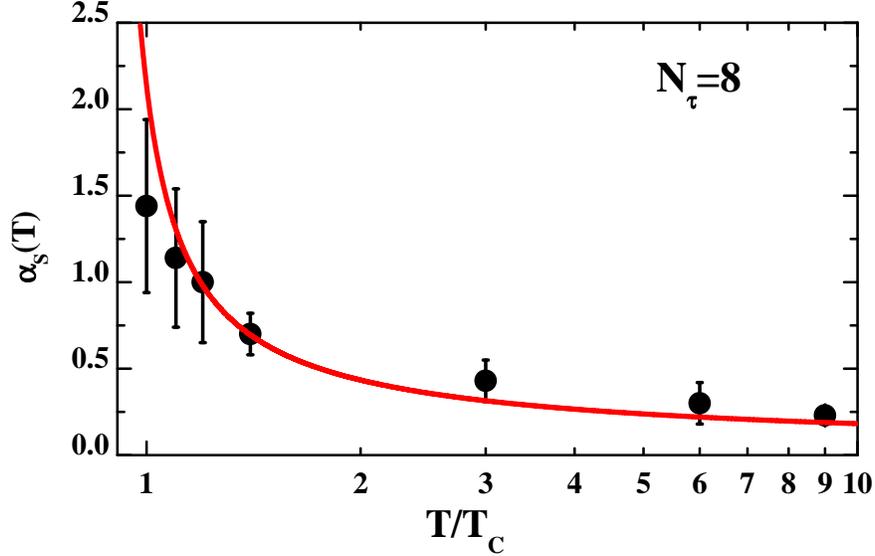}
    \caption{The coupling $\alpha_s(T) = g^2(T)/(4\pi)$ (solid red
    line) as a function of $T/T_c$ in comparison to the long range part of the
strong coupling as extracted from Ref. \protect\cite{Bielefeld} from the free energy
of a quark-antiquark pair in quenched lQCD (for $N_\tau$ = 8).  }
    \label{fig_1}
  \end{center}
\end{figure}

Following  \cite{pQP} the quasiparticle mass (squared) is written
in (momentum-independent) perturbative form,
\be
 M^2(T) = \frac{N_c}6\, g^2 T^2 \, ,
 \label{eq:M2}
\ee with a running coupling (squared),
\be
 g^2(T/T_c) = \frac{48\pi^2}{11N_c
 \ln(\lambda^2(T/T_c-T_s/T_c)^2}\ ,
 \label{eq:g2}
\ee which permits for an enhancement near $T_c$
\cite{pQP,Rafelski}. It will be shown below that an infrared
enhancement of the coupling - as also found in the lQCD
calculations in Ref. \cite{Bielefeld} for the long range part of
the $q - \bar{q}$ potential - is directly linked to the gluon
fusion/clustering scenario. In order to quantify this statement
the coupling $\alpha_s(T) = g^2(T)/(4\pi)$ is shown in Fig.
\ref{fig_1} as a function of $T/T_c$ in comparison to the long
range part of the strong coupling as extracted from Ref.
\cite{Bielefeld} from the free energy of a quark-antiquark pair in
quenched lQCD. For this comparison the actual parameters $\lambda
= 2.42$, $T_s/T_c= 0.46$  have been adopted as in Ref.
\cite{Andre}. The parametrization (\ref{eq:g2}) is seen to follow
the lQCD results - also indicating a strong enhancement close to
$T_c$ - as a function of temperature reasonably well. One should
recall that any extraction of coupling constants $\alpha_s(T)$
from lQCD is model dependent and deviations from (or agreement
with) lattice 'data' have to be considered with care. The argument
here is that the specific 'parametric form' of Eq. (\ref{eq:g2})
is not in conflict with lQCD and that the coupling $\alpha_s$ and
consequently the quasiparticle mass $M(T)$ has the right order of
magnitude.

%%%%%%%%%%%%%%%%%%%%%%%%%%%%%%%%%%%%%%%%%%%%%%%%%%%%%%%% Fig.1%%%%%%%%
%%%%%%%%%%%%%%%%%%%%%%%%%%%%%%%%%%%%%%%%%%%%%%%%%%%%%%%%%%%%%%%%%%%%%%%

The width $\gamma$ is adopted in the form $\gamma \sim g^2 T \ln
g^{-1}$ \cite{Pisar89LebedS} or, equivalently, in terms of $M$
\cite{Andre04}, as
\be
  \gamma(T)
  =
  \frac3{4\pi}\, \frac{M^2(T)}{T^2} \, T \ln\frac{c}{(M(T)/T)^2} \, ,
 \label{eq:gamma}
\ee where $c=14.4$ (from \cite{Andre}) is related to a magnetic
cut-off. In case of the pure Yang-Mills sector of QCD the physical
processes contributing to the width $\gamma$ are  both $gg
\leftrightarrow gg$ scattering as well as splitting and fusion
reactions $gg \leftrightarrow g$ or $gg \leftrightarrow ggg$, $ggg
\leftrightarrow gggg$ etc. Note that the ratio $\gamma(T)/M(T)
\sim g \ln(c/g^2)$ approaches zero only asymptotically for $T
\rightarrow \infty$ such that the width of the quasiparticles is
comparable to the mass for all practical energy scales on earth;
the ratio $\gamma(T)/M(T)$ drops below 0.5 only for temperatures
$T > 1.25\cdot 10^5 \ T_c$ (for the parameters given above).

For the choice (\ref{eq:rho}) for the spectral function the scalar
effective propagator reads, \be \Delta^{dqp}(\omega, {\bf p}) =
\frac{1}{\omega^2 - {\bf p}^2 - M^2 + 2i\gamma \omega} \ , \ee
which can easily be separated into real and imaginary parts. The
entropy density (\ref{sdqp}) then reads explicitly \cite{Andre05}, $$
s^{dqp}(T) = d_g \int  \frac{d^3 p}{(2 \pi)^3} \ \left(
-\ln(1-e^{-\omega_p/T}) + \frac{\omega_p}{T} n(\omega_p/T) \right)
 $$\be \label{sss} \hspace{1.2cm} + d_g \int \frac{d \omega}{2 \pi} \frac{d^3
p}{(2 \pi)^3} \
 \frac{\partial n}{\partial T} \left(\arctan(\frac{2 \gamma
 \omega}{\omega_p^2 - \omega^2}) - \frac{2 \gamma \omega
 (\omega_p^2 - \omega^2)}{(\omega_p^2-\omega^2)^2 + 4 \gamma^2
 \omega^2} \right) \ , \ee

\noindent
 using $\omega_p = \sqrt{{\bf p}^2 + M^2}$. The first line in (\ref{sss})
 corresponds to the familiar on-shell quasiparticle contribution $s_0$
 while the second line in (\ref{sss})  corresponds to the
 contribution originating from the finite width $\gamma$ of the
 quasiparticles and is positive throughout but subleading (see below).

The pressure $P$ now can be evaluated from
\be
\label{pressure} s =\frac{\partial P}{dT} \ee by integration of
$s$ over $T$, where  from now on we identify the 'full' entropy
density $s$ with the quasiparticle entropy density $s^{dqp}$. Note that for $T <
T_c$ the entropy density drops to zero (with decreasing $T$) due to the
high quasiparticle mass and the width $\gamma$ vanishes as well
because the interaction rate in the very dilute quasiparticle
system becomes negligible. Since the pressure for infinitely heavy
(noninteracting) particles also vanishes the integration constant
for the pressure $P$ - when integrating (\ref{pressure}) - may
safely be assumed to be zero, too.

The energy density $\epsilon$ then follows from the
thermodynamical relation \cite{pQP,Peshi} \be \label{eps} \epsilon
= T s -P \ee and thus is also fixed by the entropy $s(T)$ as well
as the interaction measure \be \label{wint} W(T): = \epsilon(T) -
3P(T) = Ts - 4 P \ee that vanishes for massless and noninteracting
degrees of freedom.

In Ref. \cite{Andre} a detailed comparison has been presented with
the lattice results from Ref. \cite{CCPACS} for the pure gluonic
sector to the quasiparticle entropy density (\ref{sss}) for the parameters
given above.  The agreement with the lattice data is practically perfect
\cite{Andre,Andre04}. Needless to point out that also $P(T),
\epsilon(T)$ and $W(T)$ well match the lattice QCD results for 1
$\leq T/T_c \leq 4$ \cite{Andre,Andre05} due to thermodynamical
consistency. The same parameters are also adopted for the following
calculations.

\subsection{Time-like and space-like quantities}
For the further argumentation  it is useful to introduce the
shorthand notation \be \label{conv}
 {\rm \tilde Tr}_P^{\pm} \cdots
 =
 d_g\!\int\!\!\frac{d \omega}{2 \pi} \frac{d^3p}{(2 \pi)^3}\,
 2\omega\, \rho(\omega)\, \Theta(\omega) \, n(\omega/T) \ \Theta(\pm P^2) \, \cdots \,
\ee with $P^2= \omega^2-{\bf p}^2$ denoting the invariant mass
squared. The $\Theta(\pm P^2)$ function in (\ref{conv}) separates
time-like quantities from space-like quantities and can be
inserted for any observable of interest.

\begin{figure}[htb!]
  \begin{center}
  \vspace{0.9cm}
    \includegraphics[width=11.5cm]{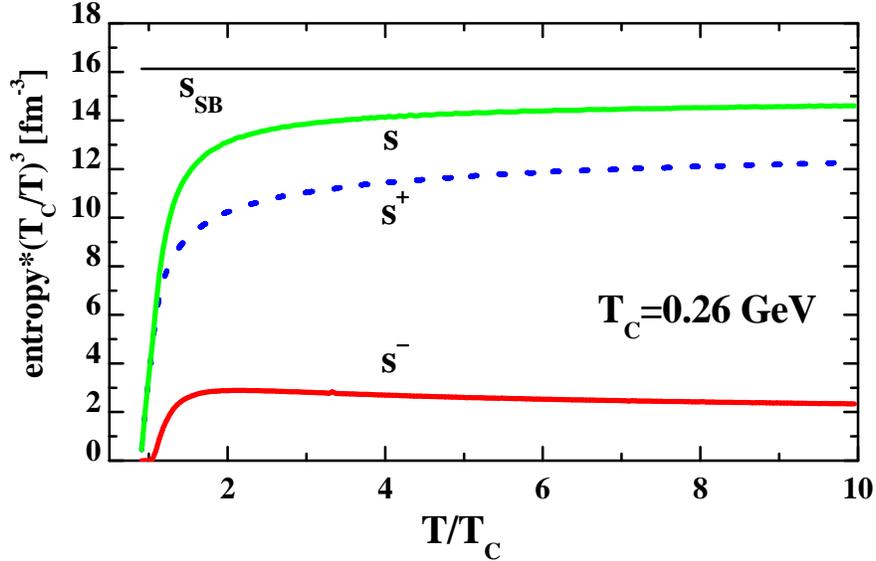}
    \caption{The time-like contribution to the entropy density
$s^+$ (dotted blue line), the space-like contribution  $s^-$
(lower red line) and the  total entropy density $s= s^+ + s^-$
(thick solid green line) as a function of $T/T_c$. All quantities
have been multiplied by the dimensionless factor $(T_c/T)^3)$
assuming $T_c$ = 0.26 GeV for the pure gluonic system
\protect\cite{Dosch}. The upper solid black line displays the
Stefan Boltzmann limit $s_{SB}$ for reference. }
    \label{fig0}
  \end{center}
\end{figure}

As the first quantity we consider the entropy density (\ref{sss}).
Its time-like contribution is almost completely dominated by the
first line in (\ref{sss}) - that corresponds to the  on-shell
quasiparticle contribution $s_0$ - but also includes a
 small contribution from the second line in (\ref{sss}) which is
 positive for $T$ below about 1.5 $T_c$ and becomes negative for
 larger temperature. This time-like part $s^+$ is shown
 in Fig. \ref{fig0} by the dotted blue line (multiplied by $(T_c/T)^3$).
 The second line in (\ref{sss}) - as mentioned above -  corresponds to the
 contribution originating from the finite width $\gamma$ of the
 quasiparticles and also has a space-like part $s^-$ which is dominant
 (for the second line in (\ref{sss})) and displayed
 in  Fig. \ref{fig0} by the lower red line (multiplied by
 $(T_c/T)^3$). Though $s^-$ is subleading
 in  the total entropy density $s = s^+ + s^-$  (thick solid green line in
 Fig. \ref{fig0}) it is
 essential for a proper reproduction of $s(T)$ close to $T_c$ (cf. \cite{Andre05}).
 Note that the total entropy density $s$ is not very different from the
 Stefan Boltzmann entropy density $s_{SB}$ for $T > 2 T_c$ as shown
 in Fig. \ref{fig0} by the
 upper thin line (multiplied by $(T_c/T)^3$).

%%%%%%%%%%%%%%%%%%%%%%%%%%%%%%%%%%%%%%%%%%%%%%%%%%%%%%%% Fig.1%%%%%%%%
%%%%%%%%%%%%%%%%%%%%%%%%%%%%%%%%%%%%%%%%%%%%%%%%%%%%%%%%%%%%%%%%%%%%%%%

Further quantities of interest are the quasiparticle 'densities'
\be
   N^\pm (T) = {\rm {\tilde Tr^\pm }}\ 1
   \label{eq: N+}
\ee that correspond to the time-like (+) and space-like (-) parts
of the integrated distribution function. Note that only the
integral of $N^+$ over space has a particle number interpretation.
In QED this corresponds to time-like photons ($\gamma^*$) which
are virtuell in intermediate processes but can also be seen
asymptotically by dileptons (e.g. $e^+ e^-$ pairs) due to the
decay $\gamma^* \rightarrow e^+e^-$ \cite{Cass99}.

A scalar density $N_s$, which is only defined in the time-like
sector,  is given by \be \label{scalar} N_s(T) = {\rm {\tilde Tr^+
}}\ \left( \frac{\sqrt{P^2}}{\omega} \right) \,  \ee and has the
virtue of being Lorentz invariant. Moreover, a scalar density can
easily be computed in transport approaches for bosons and fermions
\cite{Cass99,excita} which is of relevance for the argumentation
in Section 3.

%%%%%%%%%%%%%%%%%%%%%%%%%%%%%%%%%%%%%%%%%%%%%%%%%%%%%%%% Fig.1%%%%%%%%
\begin{figure}[htb!]
  \begin{center}
  \vspace{0.1cm}
    \includegraphics[width=11.5cm]{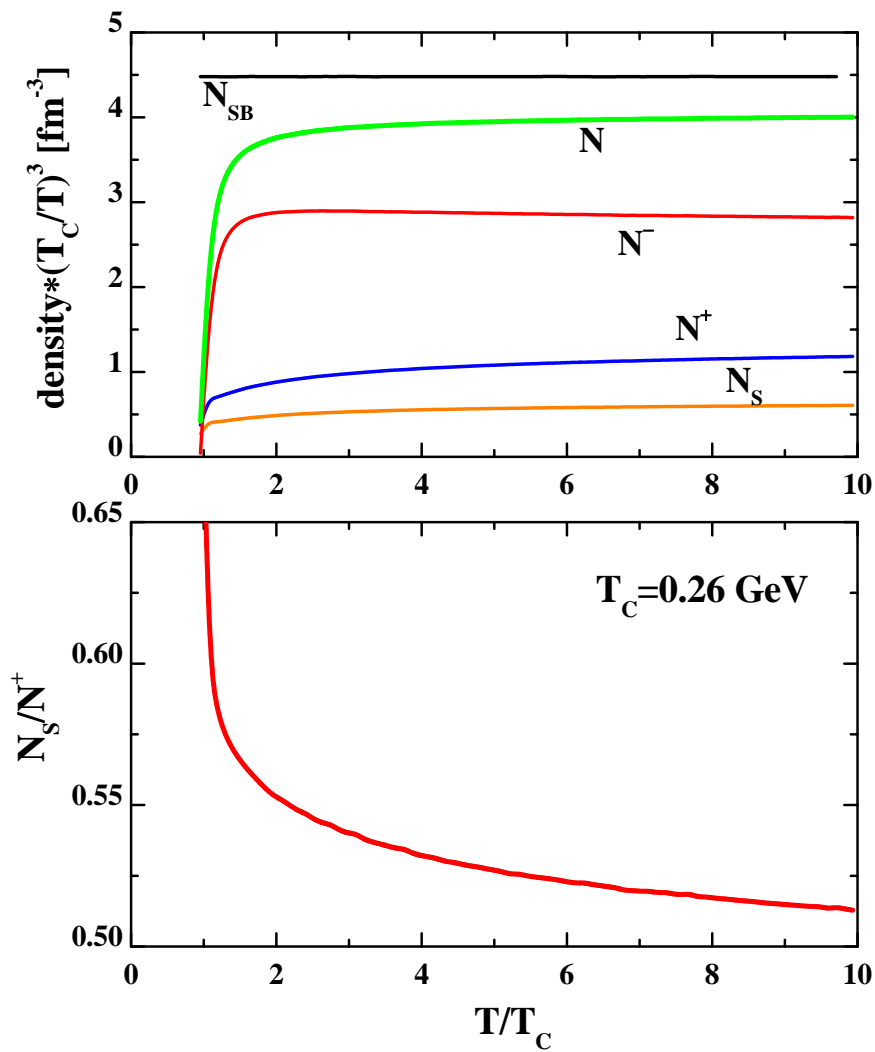}
    \caption{Upper part: The scalar density $N_s$ (lower orange line),
    the time-like density
$N^+$ (blue line), the space-like quantity $N^-$ (red line) and the
sum $N=N^+ + N^-$ (thick solid green line) as a function
of $T/T_c$ assuming $T_c$ = 0.26 GeV for the pure gluonic system
\protect\cite{Dosch}. The upper solid black line displays the
Stefan Boltzmann limit $N_{SB}$ for reference. All quantities are
multiplied by the dimensionless factor $(T_c/T)^3$. Lower part:
The ratio of the scalar density $N_s$ to the time-like density
$N^+$ as a function of the scaled temperature $T/T_c$. }
    \label{fig1}
  \end{center}
\end{figure}
%%%%%%%%%%%%%%%%%%%%%%%%%%%%%%%%%%%%%%%%%%%%%%%%%%%%%%%%%%%%%%%%%%%%%%%

The actual results for the different 'densities' (multiplied by
$(T_c/T)^3$) are displayed in the upper part of Fig. \ref{fig1}
where the lower orange line represents the scalar density $N_s$,
the blue line the time-like density $N^+$, the red line the
space-like quantity $N^-$ and the thick solid green line the sum
$N=N^+ + N^-$ as a function of $T/T_c$ assuming (as
before) $T_c$ = 0.26 GeV for the pure gluonic system \cite{Dosch}.
It is seen that $N^+$ is substantially smaller than $N^-$ in the
whole temperature range up to 10 $T_c$ where it is tacitly assumed
that the DQPM also represents lQCD results for $T
> 4 T_c$, which is not proven explicitly, but might be expected
due to the proper weak coupling limit of (\ref{eq:M2}),
(\ref{eq:gamma}) (cf. Fig. 1). The application of the DQPM to 10
$T_c$ is presented in Fig. \ref{fig1} since the initial state at
Large Hadron Collider (LHC) energies might be characterized by a
temperature above 4 $T_c$; note that the properties of the
partonic phase will be explored  from the experimental side in the
near future at LHC. Quite remarkably the quantity $N$
follows closely the Stefan Boltzmann limit $N_{SB}$ for a massless
noninteracting system which is given in Fig. \ref{fig1} by the
upper thin solid line and has the physical interpretation of a gluon
density. Though $N$ differs
by less than 15\% from the Stefan Boltzmann (SB) limit for $T > 2
T_C$ the physical interpretation is essentially different! Whereas
in the SB limit all gluons move on the light cone without
interactions only a small fraction of gluons can be attributed to
quasiparticles with density $N^+$ within the DQPM that propagate
within the lightcone. The space-like part $N^-$ corresponds to
'gluons' exchanged in $t$-channel scattering processes and thus
cannot be propagated explicitly in off-shell transport approaches
without violating causality and/or Lorentz invariance.

The scalar density $N_s$ follows smoothly the time-like density
$N^+$ as a function of temperature which can be explicitly seen in
the lower part of Fig. \ref{fig1} where the ratio $N_s/N^+$ is
shown versus $T/T_c$. Consequently, the scalar density $N_s$
uniquely relates to the time-like density $N^+$ or the temperature
$T$ in thermal equilibrium which will provide some perspectives
for a transport theoretical treatment (see Section 3).

The separation of $N^+$ and $N^-$ so far has no direct dynamical
implications except for the fact that only the fraction $N^+$ can
explicitly be propagated in transport as argued above. Thus we
consider the energy densities,
 \be \label{energy} T_{00}^\pm(T) = {\rm {\tilde Tr^\pm
}}\ \omega  \ , \ee
that specify time-like and space-like contributions
to the quasiparticle energy density. It is worth pointing out that
the quantity $T_{00} = T_{00}^+ + T_{00}^-$ in case of a
conventional quasi-particle model with vanishing width $\gamma$ in
general is quite different from $\epsilon$ in (\ref{eps}) because
the interaction energy density in this case is not included in
(\ref{energy}), i.e. \be \label{qqp} T_{00} = T_{00}^+ =
d_g\!\int\!\! {d \omega} \frac{d^3p}{(2 \pi)^3}\,
 2\omega\, \delta(\omega^2-M^2-{\bf p}^2)\, \Theta(\omega) \, \Theta(\pm P^2)\,
 n(\omega/T) \ \omega \,  \ee
since $\omega^2-{\bf p}^2 = M^2 = P^2 > 0$ due to the mass-shell
$\delta$-function.

\begin{figure}[htb!]
%  \begin{center}
    \includegraphics[width=11.5cm]{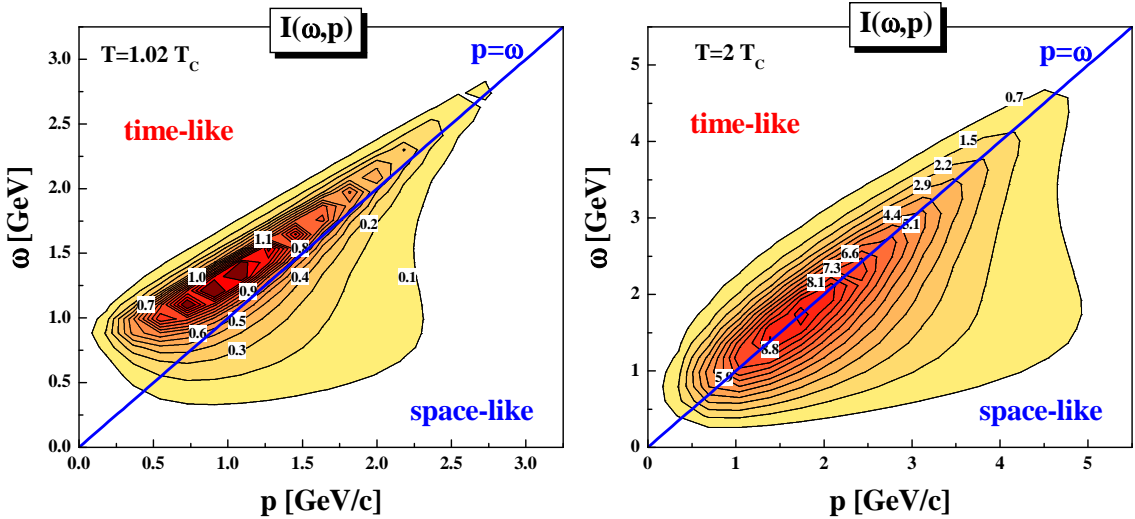}
    \caption{The integrand $I(\omega, p)$ (\ref{explain}) for $T=1.02 T_c$ (l.h.s.)
    and $T=2 T_c$ (r.h.s.) in terms of contour lines. The straight (blue)
    line ($\omega = p$)
    separates the lime-like from the space-like sector.  Note that for a convergence
    of the energy density integral the upper limits for
    $\omega$ and $p$ have to be increased by
    roughly an order of magnitude compared to the area shown in the figure. }
    \label{fignew}
%  \end{center}
\end{figure}

How does the situation look like in case of dynamical
quasiparticles of finite width? To this aim we consider
the integrand in the energy density (\ref{energy}) which reads as
(in spherical momentum coordinates with angular
degrees of freedom integrated out)
\be
\label{explain}
I(\omega, p) =  \frac{d_g}{2 \pi^3}\ p^2 \ \omega^2
\, \rho(\omega,p^2)\, n(\omega/T)  \, .
\ee
Here the integration is to be taken over $\omega$ and $p$ from $0$
to $\infty$. The integrand $I(\omega, p)$ is shown in Fig.
\ref{fignew} for $T=1.02 T_c$ (l.h.s.) and $T=2 T_c$ (r.h.s.) in
terms of contour lines. For the lower temperature the gluon mass
is about 0.91 GeV and the width $\gamma \approx $ 0.15 GeV such that
the quasiparticle properties are close to a $\rho$-meson in free
space. In this case the integrand $I(\omega,p)$ is essentially
located in the time-like sector and the integral over the
space-like sector is subdominant. This situation changes for $T =
2 T_c$ where the mass is about 0.86 GeV while the width increases to
$\gamma \approx $ 0.56 GeV. As one observes from the r.h.s. of
Fig. \ref{fignew} the maximum of the integrand is shifted towards
the line $\omega = p$ and higher momentum due to the increase
in temperature by about a factor of two; furthermore,
the distribution reaches far out in the
space-like sector due to the Bose factor $n(\omega/T)$ which
favors small $\omega$. Thus the
relative importance of the time-like (+) part to the space-like (-) part is
dominantly controlled by the width $\gamma$ - relative to the pole mass -
which determines the fraction
of $T_{00}^-$ with negative invariant mass squared $(P^2 < 0)$ relative to
the time-like part $T_{00}^+$.

\vspace{1cm}
%%%%%%%%%%%%%%%%%%%%%%%%%%%%%%%%%%%%%%%%%%%%%%%%%%%%%%%% Fig.1%%%%%%%%
%%%%%%%%%%%%%%%%%%%%%%%%%%%%%%%%%%%%%%%%%%%%%%%%%%%%%%%%%%%%%%%%%%%%%%%

%%%%%%%%%%%%%%%%%%%%%%%%%%%%%%%%%%%%%%%%%%%%%%%%%%%%%%%% Fig.1%%%%%%%%
\begin{figure}[htb!]
  \begin{center}
    \includegraphics[width=11.5cm]{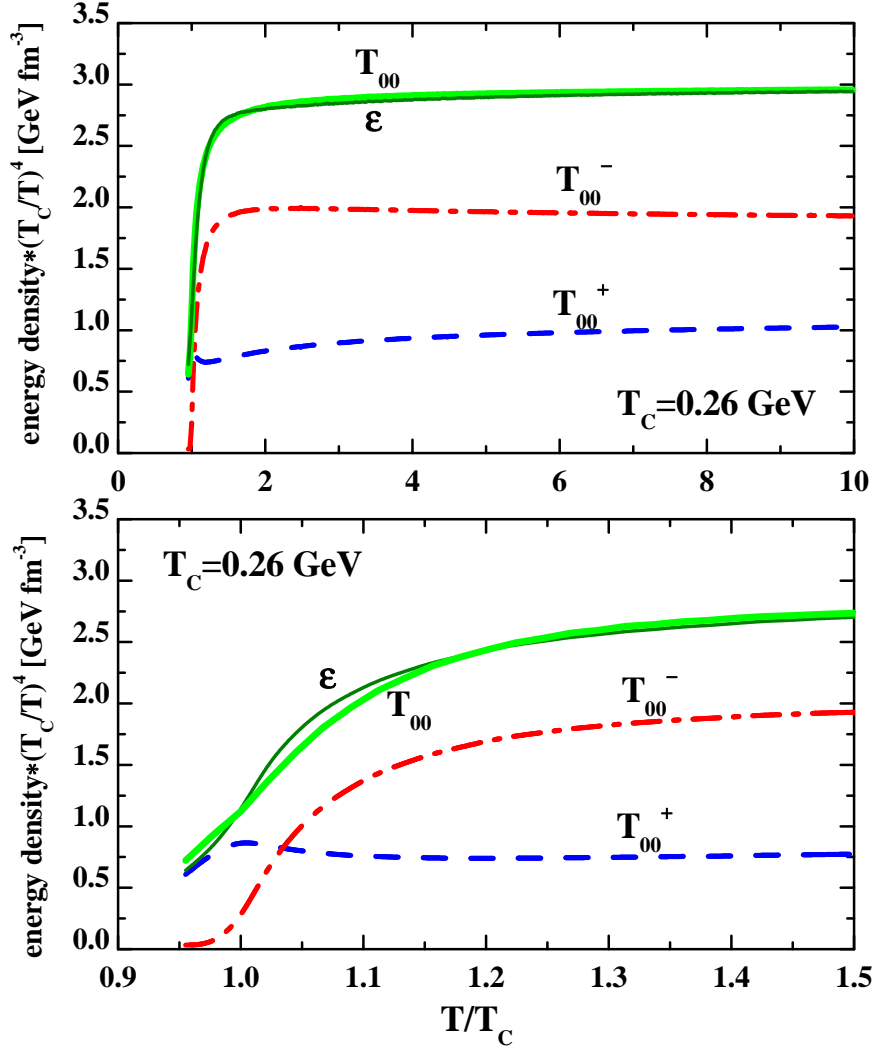}
    \caption{Upper part: The time-like energy density $T_{00}^+$ (dashed blue line),
 the space-like energy density $T_{00}^-$  (dot-dashed red line)
and the total energy density $T_{00}=T_{00}^+ + T_{00}^-$ (thick
solid green line) as a function of $T/T_c$. The thin black line
displays the energy density $\epsilon(T/T_c)$ from
(\protect\ref{eps}); it practically coincides with $T_{00}$ within
the linewidth and is hardly visible. All densities are multiplied
by the dimensionless factor $(T_c/T)^4$ in order to divide out the
leading temperature dependence. Lower part: Same as the upper part
in order to enhance the resolution close to $T_c$.
   }
    \label{fig2}
  \end{center}
\end{figure}
%%%%%%%%%%%%%%%%%%%%%%%%%%%%%%%%%%%%%%%%%%%%%%%%%%%%%%%%%%%%%%%%%%%%%%%

The explicit results for the quasiparticle energy densities $T^+_{00}$ and
$T^-_{00}$ are displayed in Fig. \ref{fig2} by the dashed blue and
dot-dashed red lines (multiplied by $(T_C/T)^4$), respectively. As
in case of  $N^+$ and $N^-$ the space-like energy
density $T_{00}^-$ is seen to be larger than the time-like
part $T_{00}^+$ for all temperatures above 1.05 $T_c$. Since the
time-like part $T^+_{00}$ corresponds to the independent
quasiparticle energy density within the lightcone, the space-like
part $T^-_{00}$ can be interpreted as an interaction density $V$
if the quasiparticle energy $T_{00}$ matches the total energy
density $\epsilon(T)$ (\ref{eps}) as determined from the
thermodynamical relations (\ref{pressure}) and (\ref{eps}). In
fact, the DQPM yields an energy density $T_{00}$ - adding up the
space-like and time-like parts - that almost coincides with
$\epsilon(T)$ from (\ref{eps}) as seen in Fig. \ref{fig2} where
both quantities (multiplied by $(T_C/T)^4$) are displayed in terms
of the thin black and thick solid green lines, respectively;
actually both results practically coincide within the linewidth
for $T> 2 T_c$. An explicit representation of their numerical
ratio gives unity within 2\% for $T> 2 T_c$; the remaining
differences can be attributed to temperature derivatives $\sim
d/dT (\ln (\gamma/E))$ etc. in order to achieve thermodynamic
consistency but this is not the primary issue here and will be
discussed in a forthcoming study \cite{Cass07}. The deviations are
more clearly visible close to $T_c$ (lower part of Fig. 2) where
the variation of the width and mass are most pronounced. However,
for all practical purposes one may consider $T_{00}(T) \approx
\epsilon(T)$ and separate the kinetic energy density $T^+_{00}$
from the potential energy density $T^-_{00}$ as a function of $T$
or - in equilibrium - as a function of the scalar gluon density
$N_s$ or $N^+$, respectively.

\section{Dynamics of time-like quasiparticles}
Since in transport dynamical approaches there are no
thermodynamical Lagrange parameters like the inverse temperature
$\beta = T^{-1}$ or the quark chemical potential $\mu_q$, which
have to be introduced in thermodynamics in order to specify the
average values of conserved quantities (or currents in the
relativistic sense), derivatives of physical quantities with
respect to the scalar density $\rho_s = N_s$  (or time-like gluon
density $\rho_g = N^+$) are considered in the following (cf. Ref.
\cite{Toneev}). As mentioned above one may relate derivatives in
thermodynamic equilibrium via,
\be \label{DT} \frac{d}{dT} =
\frac{d}{d \rho_s} \ \frac{d \rho_s}{d T}, \ee
if the volume and
pressure are kept constant. For example, a numerical evaluation of
$d \rho_s/d (T/T_c)$ gives
\be
\label{fit1} \frac{d \rho_s}{d (T/T_c)} \approx a_1
\left(\frac{T}{T_c} \right)^{2.1} - a_2 \exp(-b(\frac{T}{T_c})) \,
\ee with $b$= 5, $a_1 = 1.5 fm^{-3}$ and $a_2 = 104 fm^{-3}$,
which follows closely the quadratic scaling in $T/T_c$ as expected
in the Stefan Boltzmann limit. The additional exponential term in
(\ref{fit1}) provides a sizeable correction close to $T_c$. The
approximation (\ref{fit1}) may be exploited for convenient
conversions between $\rho_s$ and $T/T_c$ in the pure gluon case
but will not be explicitly used in the following.

The independent quasiparticle energy density $T_K:= T_{00}^+$ and
potential energy density  $V : = T_{00}^-$ now may be expressed as
functions of $\rho_s$ (or $\rho_g$) instead of the temperature
$T$. The interaction energy density then might be considered as a scalar
energy density which - as in the nonlinear $\sigma$-model for
baryonic matter \cite{SIGMAM} - is a nonlinear function of the
scalar density $\rho_s$. As in case of nuclear matter problems the
scalar density $\rho_s$ does not correspond to a conserved
quantity when integrating over space; it only specifies the
interaction density parametrically, i.e. $V(\rho_s)$.
Alternatively one might separate $V$ into parts with different
Lorentz structure, e.g. scalar and vector parts as in case of
nuclear matter problems \cite{SIGMAM}, but this requires additional information
that cannot be deduced from the DQPM alone.

\begin{figure}[htb!]
  \begin{center}
    \includegraphics[width=11.5cm]{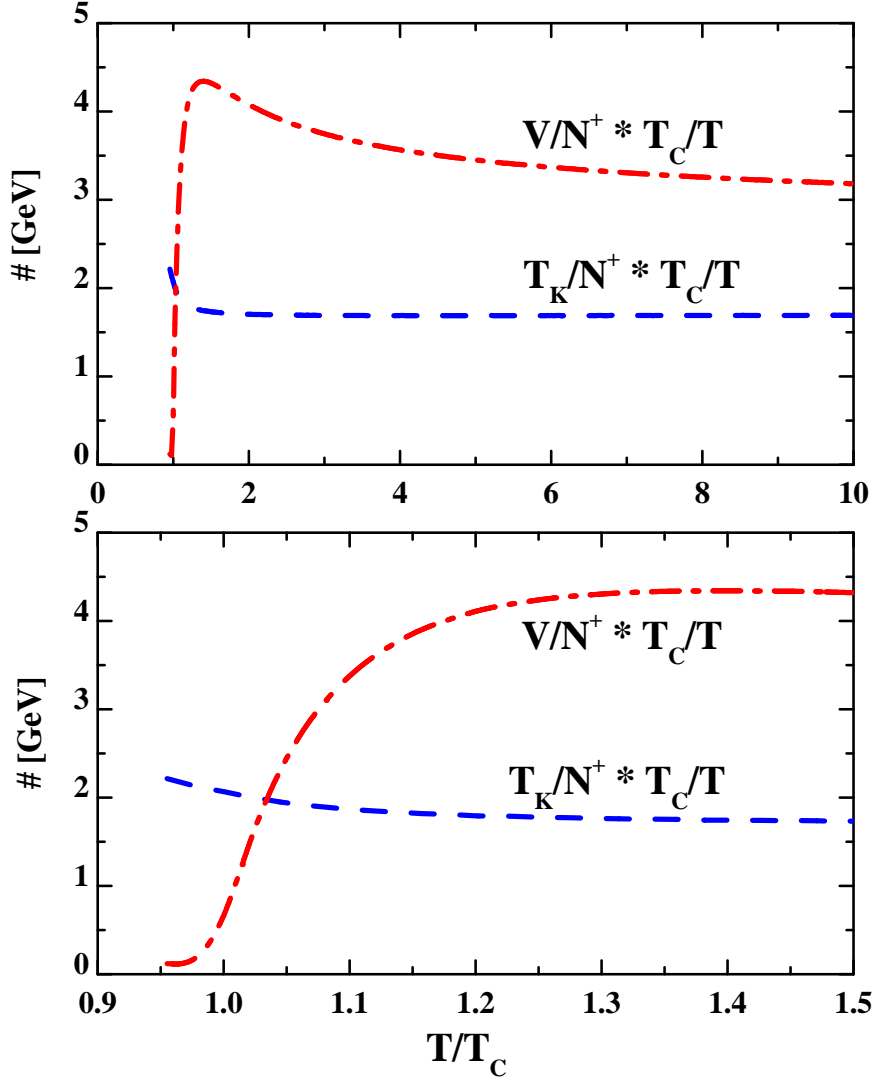}
    \caption{Upper part: The quasiparticle  energy per degree of freedom $T_K/N^+$
    (dashed blue line) and  the space-like potential energy per degree of freedom
     $V/N^+$  (dot-dashed red line) as a function of $T/T_c$.
All energies are multiplied by the dimensionless factor $(T_c/T)$.
Lower part: Same as the upper part in order to enhance the
resolution close to $T_c$.
   }
    \label{fig3}
  \end{center}
\end{figure}

It is instructive to show the 'quasiparticle' and potential energy
per degree of freedom $T_K/N^+$ and $V/N^+$ as a function of e.g.
$N^+$, $N_s$ or $T/T_c$. As one might have anti\-cipated the
kinetic energy  per effective degree of freedom is smaller than
the respective potential energy for $T/T_c > $ 1.05 as seen from
Fig. \ref{fig3} where both quantities are displayed as a function
of $T/T_c$ in terms of the dashed and dot-dashed line,
respectively. It is seen that the potential energy per degree of
freedom steeply rises in the vicinity of $T_c$ whereas the
independent quasiparticle energy rises almost linearly with $T$.
Consequently rapid changes in the
 density - as in the expansion of the fireball in
ultrarelativistic nucleus-nucleus collisions - are accompanied by
a dramatic change in the potential energy density and thus to a
violent acceleration of the quasi-particles. It is speculated here
that the large collective flow of practically all hadrons seen at
RHIC \cite{STARS} might be attributed to the early strong partonic
forces expected from the DQPM.

\begin{figure}[htb!]
  \begin{center}
  \vspace{0.8cm}
    \includegraphics[width=11.5cm]{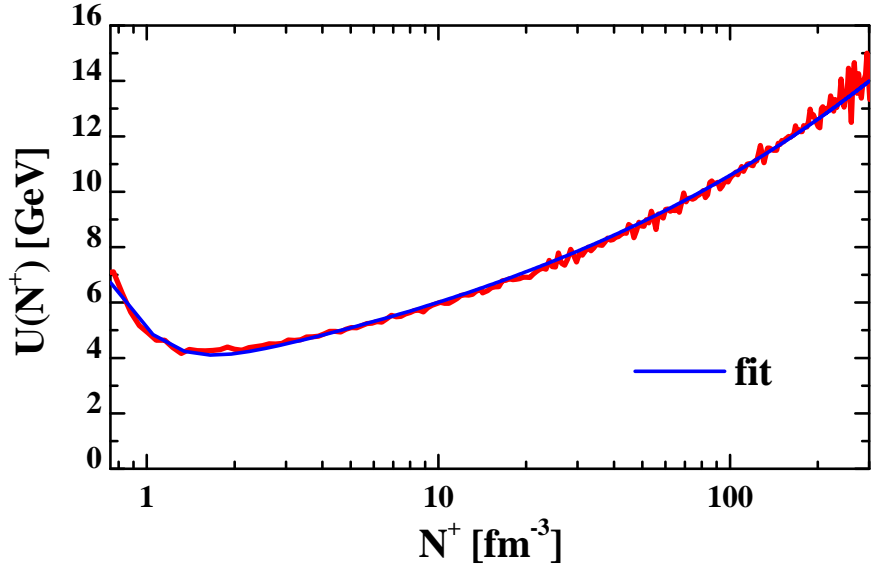}
    \caption{The mean-field potential $U(N^+)= U(\rho_g)$
     as a function of the time-like gluon density $N^+ =\rho_g$
    in comparison to the fit (\ref{pott}) (solid blue line). The densities $N^+$= 1, 1.4,
    5, 10, 50, 100 fm$^{-3}$ correspond to scaled temperatures of $T/T_c
    \approx$ 1.025, 1.045, 1.25, 1.5, 2.58, 3.25, respectively (cf. Fig. \ref{fig1}).}
    \label{fig4}
  \end{center}
\end{figure}

In order to obtain some idea about the mean-field potential
$U_s(\rho_s)$ (or $U(\rho_g)$ in the rest frame) one can consider
the derivative $d V/\rho_s = U_s(\rho_s)$ or $d V/N^+ = U(N^+) =
U(\rho_g) $. The latter is displayed in Fig. \ref{fig4} as a
function of $N^+= \rho_g$ and shows a distinct minimum at $\rho_g
\approx$ 1.4 fm$^{-3}$ which corresponds to a temperature $T
\approx 1.045 T_c$. The actual numerical results can be fitted by
the expression,
\be
\label{pott} U(\rho_g) = \frac{d V}{d \rho_g} \approx 39 \
e^{-\rho_g/0.31} + 2.93\  \rho_g^{0.21} + 0.55\  \rho_g^{0.36} \,
\ \ [{\rm GeV}] \ , \ee where $\rho_g$ is given in fm$^{-3}$ and
the actual numbers in front carry a dimension in order to match to
the proper units of GeV for the mean-field $U$. By analytical
integration of (\ref{pott}) one obtains a suitable approximation
to $V(\rho_g)$. The approximation (\ref{pott}) works sufficiently
well as can be seen from Fig. \ref{fig4} - showing a comparison of
the numerical derivative $d V/d N^+$ with the fit (\ref{pott}) in
the interval 0.7 fm$^{-3} < N^+ \leq $ 300 fm$^{-3}$ - such that
one may even proceed with further analytical calculations. Note
that a conversion between the time-like quasiparticle density $N^+
=\rho_g$ and the scalar density $\rho_s$ is easily available
numerically (cf. lower part of Fig. 3) such that derivatives with
respect to $\rho_s$ are at hand, too; the latter actually enter
the explicit transport calculations \cite{PHSD} while derivatives
with respect to $\rho_g$ in the rest frame of the system are more
suitable for physical interpretation and will be used below.

Some information on the properties of the effective gluon-gluon
interaction $v_{gg}$ may be extracted from the second derivative
of $V$ with respect to $\rho_g$, i.e. \be \label{interaction}
v_{gg}(\rho_g): = \frac{d^2 V}{d \rho_g^{2}} \approx -125.8\
e^{-\rho_g/0.31} + 0.615/ \rho_g^{0.79} + 0.2/\rho_g^{0.64} \, \ \
[{\rm GeV fm^3}] , \ee where the numbers in front have again a
dimension to match the units of GeV fm$^3$. The effective
gluon-gluon interaction $v_{gg}$ (\ref{interaction}) is strongly
attractive at low density 0.003 fm$^{-3} < \rho_g$ and changes
sign at $\rho_g \approx$ 1.4 fm$^{-3}$ to become repulsive at
higher densities. Note that the change of quasiparticle momenta
(apart from collisions) will be essentially driven by the
(negative) space-derivatives $-\nabla U(x) = - d U(\rho_g)/d
\rho_g \ \nabla \rho_g(x)$ (or alternatively by $- d U_s(\rho_s)/d
\rho_s \ \nabla \rho_s(x)$). This implies that the gluonic
quasiparticles (at low gluon density) will bind with decreasing
density, i.e. form 'glueballs' dynamically close to the phase
boundary and repell each other for $\rho_g \geq$ 1.4 fm$^{-3}$.
Note that color neutrality is imposed by color-current
conservation and only acts as a boundary condition for the quantum
numbers of the bound/resonant states in color space.

This situation is somehow reminiscent of the nuclear matter problem
\cite{SIGMAM} where a change in sign of the 2nd derivative of the
potential energy density of nuclear matter at low density
indicates the onset of clustering of nucleons, i.e. to deuterons,
tritons, $\alpha$-particles etc., which form the states of the
many-body system at low nucleon densities (and not a low density
nucleon gas). This is easy to follow up for the simplified
nonrelativistic energy density functional $\epsilon_N$ for nuclear
matter, \be \label{nmatter} \epsilon_N \approx A \rho_N^{5/3} +
\frac{B}{2} \rho_N^2 + \frac{3C}{7} \rho^{7/3}_N , \ee where the
first term gives the kinetic energy density and the second and
third term correspond to attractive and repulsive interaction
densities. For $A \approx 0.073 $GeV fm$^2$, $B \approx -1.3$ GeV
fm$^3$ and $C \approx 1.78$ GeV fm$^4$ a suitable energy density
for nuclear matter is achieved; it gives a minimum in the energy
per nucleon $E/A = \epsilon_N/\rho_N \approx - 0.016$ GeV for
nuclear saturation density $\rho_N^0 \approx 0.168$ fm$^{-3}$. The
mean-field potential $U_N = B \rho_N + C \rho_N^{4/3}$ has a
minimum close to $\rho_N^0$ such that the effective
nucleon-nucleon interaction strength $v_{NN} = B + 4/3 C
\rho_N^{1/3}$ changes from attraction to repulsion at this
density. Note that in the gluonic case the minimum in the
mean-field potential $U$ (\ref{pott}) occurs at roughly 8 times
$\rho_N^0$ and the strength of the gluonic interaction is higher
by more than 2 orders of magnitude!

The confining nature of the effective gluon-gluon interaction
$v_{gg}$ (\ref{interaction}) becomes apparent in the limit $\rho_g
\rightarrow 0$, where the huge negative exponential term dominates
for $\rho_g >$ 0.003 fm$^{-3}$; for even smaller densities the
singular repulsive terms take over. Note, however, that the
functional extrapolation of the fit (\ref{pott}) to vanishing
gluon density $\rho_g$ has to be considered with care and it
should only be concluded that the interaction strength becomes
'very large'. On the other hand the limit $\rho_g \rightarrow 0$
is only academical because the condensation/fusion dynamically occurs
for $\rho_g \approx$ 1 fm$^{-3}$.

A straight forward way to model the gluon condensation or
clustering to confined glueballs dynamically (close to the phase transition)
is to adopt a screened Coulomb-like potential $v_c(r,\Lambda)$
with the strength $\int d^3r \ v_c(r,\Lambda)$ fixed by
$v_{gg}(\rho_g)$ from (\ref{interaction}) and the screening length
$\Lambda$ from lQCD studies. For the 'dilute gluon regime'
($\rho_g < $ 1.4 fm$^{-3}$), where two-body interactions should dominate,
 one may solve a Schr\"odinger (or
Klein-Gordon) equation for the bound and/or resonant states. This
task is not addressed further in the present study since for the actual
applications (as in the Parton-Hadron-String-Dynamics (PHSD)
approach \cite{PHSD}) dynamical quark and antiquarks have to be
included. The latter degrees of freedom do not change the general
picture very much for higher temperatures $T > 2 T_c$ but the
actual numbers are different close to $T_c$ since the quarks and
antiquarks here dominate over the gluons due to their lower mass.
The reader is referred to an upcoming study in Ref. \cite{Cass07}.

\begin{figure}[htb!]
  \begin{center}
    \includegraphics[width=11.5cm]{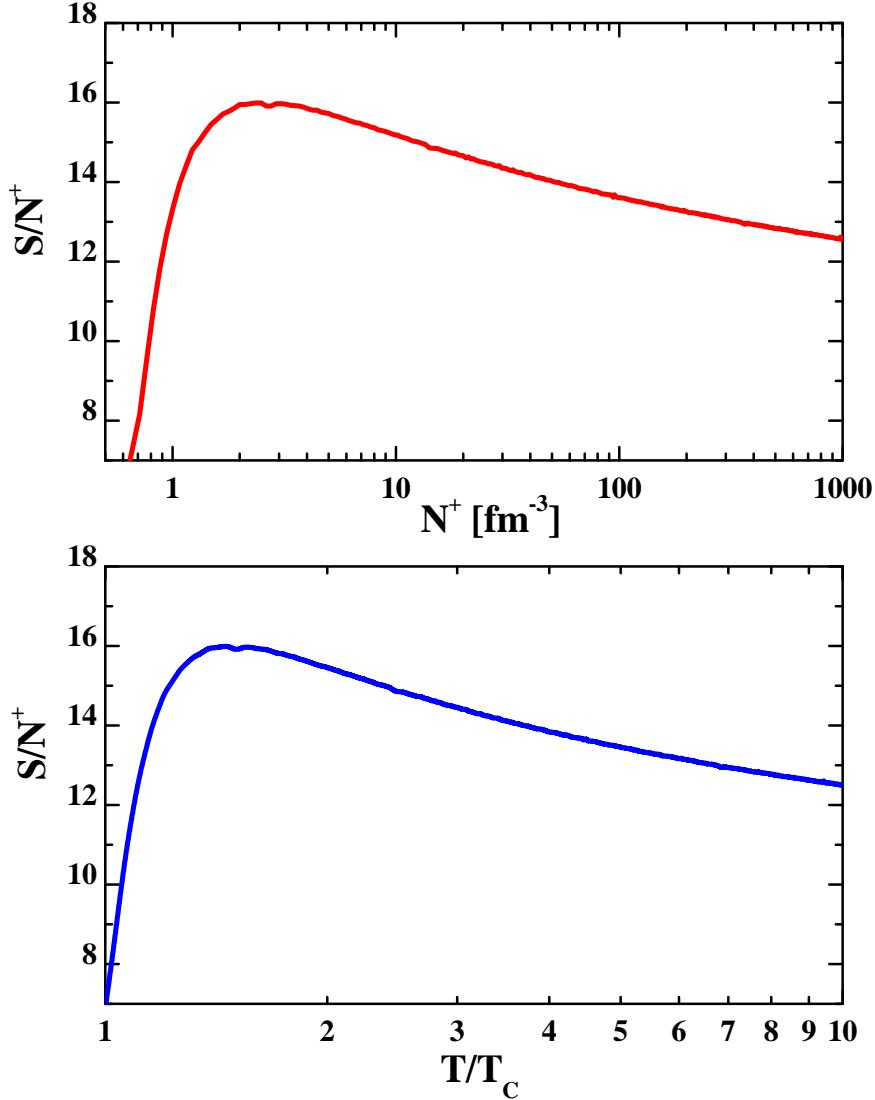}
    \caption{The entropy per degree of freedom $s/N^+$
 as a function of $N^+$ (upper part) or $T/T_c$ (lower part).}
    \label{fig5}
  \end{center}
\end{figure}

Some comments on expanding gluonic systems in equilibrium appear
in place, i.e. for processes where the total volume $\tilde{ V}$
and pressure $P$ play an additional role. For orientation we show
the entropy per time-like particle $s/N^+$ in Fig. \ref{fig5} as a
function of $N^+$ (upper) and  $T/T_c$ (lower part) which drops
close to the phase boundary since the quasiparticles become weakly
interacting (cf. Fig. \ref{fig3}). Note that this is essentially
due to the low density and not due to the interaction strength
(\ref{interaction}); a decrease of the width $\gamma$ (as encoded
in (\ref{eq:gamma})) implies a decrease in the interaction rate! An expansion
process with conserved total entropy $S= s \tilde{ V}$ leads to a
change in the total gluon number $N^+ \tilde{ V}$ since $s/N^+$
changes with density (or temperature) (Fig. \ref{fig5}). The same
holds for an expansion process with constant total energy
$\epsilon \tilde{ V}$ since also $\epsilon/N^+$ is varying with
density (or temperature). Other scenarios involving e.g. $S = P/T$
also involve a change of the gluon number $N^+ \tilde{ V}$ during
the cooling process such that reactions like $gg \leftrightarrow
g$, $ggg \leftrightarrow gg$ etc. are necessary ingredients of any
transport theoretical approximation. We do not further investigate
different expansion scenarios here since the reactions $g
\leftrightarrow q\bar{q}$, i.e. the gluon splitting to a quark and
antiquark as well as the backward fusion process, are found to play a dominant
role in the vicinity of the phase transition as well as for higher
temperatures \cite{Cass07,PHSD}.

\section{Conclusions and discussion}
The present study has provided a novel interpretation of the
dynamical quasiparticle model (DQPM) by separating time-like and
space-like quantities for 'particle densities', energy densities,
entropy densities ect. that also paves the way for an off-shell
transport approach \cite{PHSD}. The entropy density $s$ in
(\ref{sss}) is found to be dominated by the on-shell quasiparticle
contribution (first line in (\ref{sss})) (cf. \cite{Andre05})
while the space-like part of the off-shell contribution (second
line in (\ref{sss})) gives only a small (but important)
enhancement (cf. Fig. 2). However, in case of the 'gluon density' $N
= N^+ + N^-$ and the gluon energy density $T_{00} = T_{00}^+ +
T_{00}^-$ the situation is opposite: here the space-like parts
($N^-, T_{00}^-$) dominate over the time-like parts ($N^+,
T_{00}^+$) except close to $T_c$ where the independent
quasiparticle limit is approximately regained. The latter limit is
a direct consequence of the infrared enhancement of the coupling
(\ref{eq:g2}) close to $T_c$ (in line with the lQCD studies in
Ref. \cite{Bielefeld} ) and a decrease of the width $\gamma$
(\ref{eq:gamma}) when approaching $T_c$ from above.

Since only the time-like part  $N^+$ can be
propagated within the lightcone the space-like part $N^-$ has to
be attributed to $t$-channel exchange gluons in scattering
processes that contribute also to the space-like energy density
$T_{00}^-$. The latter quantity may be regarded as potential
energy density $V$. This, in fact, is legitimate since the
quasiparticle energy density $T_{00}$ very well matches the energy
density (\ref{eps}) obtained from the thermodynamical relations.
Only small deviations close to $T_c$ indicate that the DQPM in its
straightforward application is not thermodynamically consistent.
However, by accounting for 'rearrangement terms' in the energy
density - as known from the nuclear many-body problem
\cite{Lenske} - full thermodynamical consistency may be regained
\cite{Cass07}.

It is instructive to compare the present DQPM to other recent
models. In the PNJL\footnote{Polyakov-loop-extended Nambu
Jona-Lasinio} model \cite{Ratti} the gluonic pressure is build up
by a constant effective potential $U(\Phi, \Phi^*;T)$ which
controls the thermodynamics of the Polyakov loop $\Phi$. It is
expanded in powers of $\Phi \Phi^*$ with temperature dependent
coefficients in order to match lQCD thermodynamics. Thus in the
PNJL there are no time-like gluons; the effective potential
$U(\Phi, \Phi^*;T)$ stands for a static gluonic pressure that
couples to the quark/antiquark degrees of freedom. The latter are
treated in mean-field approximation, i.e. without dynamical width,
whereas the DQPM incorporates a sizeable width $\gamma$.

Another approach to model lQCD thermodynamics has been suggested
in Ref. \cite{Toneev} and is based on an effective Lagrangian
which is nonlinear in the effective quark and gluon fields. In
this way the authors avoid a parametrization of the interaction
density in terms of Lagrange parameters ($T, \mu$) and achieve
thermodynamical consistency. The latter approach is closer in
spirit to the actual interpretation of the DQPM and may be well
suited for an on-shell transport theoretical formulation. The
on-shell restriction here comes about since effective Lagrangian
approaches should only be evaluated in the mean-field limit which
implies vanishing scattering width for the quasiparticles. This is
sufficient to describe systems is thermodynamical equilibrium,
where forward and backward interaction rates are the same, but
might not provide the proper dynamics out-of-equilibrium.

Some note of caution with respect to the present DQPM appears
appropriate: the parameters in the effective coupling
(\ref{eq:g2}) and the width (\ref{eq:gamma}) have been fixed in
the DQPM by the entropy density (\ref{sss}) to lQCD results assuming the
form (\ref{eq:rho}) for the spectral function $\rho(\omega)$. Alternative
assumptions for $\rho(\omega)$ will lead to slightly different
results for the time-like density, energy
densities {\it etc.} but not to a qualitatively different picture.
Independent quantities from lQCD should allow to put further
constraints on the more precise form of $\rho(\omega)$ such as
calculations for transport coefficients \cite{lattice2};
unfortunately such lQCD studies are only at the beginning. A more
important issue is presently to extend the DQPM to incorporate
dynamical quark and antiquark degrees of freedom (as in
\cite{Andre05}) in order to catch the physics of gluon splitting
and quark-antiquark fusion ($g \leftrightarrow q+\bar{q}$, $g+g
\leftrightarrow q+\bar{q}+g$) reactions \cite{Cass07,PHSD}.

Coming back to the questions raised in the Introduction concerning i) the
appropriate description of QCD thermodynamics within the DQPM and ii) the
possibility to develop a consistent off-shell partonic transport approach
as well as iii) the perspectives for a dynamical description of the
transition from partonic to hadronic
degrees of freedom, we are now in the position to state: most likely 'Yes'.

\vspace{0.5cm} The author acknowledges valuable discussions with
E. L. Bratkovskaya and A. Peshier. Furthermore he likes to thank
S. Leupold for a critical reading of the manuscript and constructive
suggestions.

%------------------------------------------------------------------------


\begin{thebibliography}{99}
\bibitem{QM01}
    {\it Quark Matter 2002}, Nucl. Phys. A { 715} (2003) 1;
    {\it Quark Matter 2004}, J. Phys. G { 30}  (2004) S633;
    {\it Quark Matter 2005}, Nucl. Phys. A 774 (2006) 1.

\bibitem{Karsch} F. Karsch {\it et al.}, Nucl. Phys. B 502 (2001) 321.

\bibitem{Thoma} M. H. Thoma, J. Phys. G 31 (2005) L7; Nucl. Phys. A 774 (2006) 307.

\bibitem{Andre} A. Peshier and W. Cassing, Phys. Rev. Lett. 94
(2005) 172301.

\bibitem{Shuryak} E. Shuryak, Prog. Part. Nucl. Phys. 53 (2004)
273.

\bibitem{STARS} I. Arsene {\it et al.}, Nucl. Phys. A 757 (2005)
1; B. B. Back {\it et al.}, Nucl. Phys. A 757 (2005) 28; J. Adams
{\it et al.}, Nucl. Phys. A 757 (2005) 102; K. Adcox {\it et al.},
Nucl. Phys. A 757 (2005) 184.

\bibitem{Miklos3} T. Hirano and M. Gyulassy, Nucl. Phys. A 769
(2006) 71.

\bibitem{Cassing03} W. Cassing, K. Gallmeister, and C. Greiner, Nucl.~Phys.~A {735}
(2004) 277.

\bibitem{Brat04} E. L. Bratkovskaya {\it et al.}, Phys. Rev. C 67 (2003) 054905;
Phys. Rev. C 69 (2004) 054907; Phys. Rev. C 71 (2005) 044901.

\bibitem{Cassing04} K. Gallmeister and W. Cassing, Nucl. Phys. A748 (2005) 241.

\bibitem{Heinz} P. Kolb and U. Heinz, nucl-th/0305084, in 'Quark
Gluon Plasma 3', Eds. R. C. Hwa and X.-N. Wang, World Scientific,
Singapore, 2004.

\bibitem{Bass2} C. Nonaka and S. A. Bass, Phys. Rev. C 75 (2007) 014902;
Nucl. Phys. A 774 (2006) 873.

\bibitem{GerryEd} G. E. Brown, C.-H. Lee, M. Rho, and E. Shuryak,
Nucl.\ Phys.\ A740 (2004) 171.

\bibitem{GerryRho} G. E. Brown, C.-H. Lee, and M. Rho, Nucl. Phys. A 747 (2005) 530.

\bibitem{Eddi} E. V. Shuryak and I. Zahed, Phys. Rev. D 70 (2004) 054507.

\bibitem{lattice2} A. Nakamura and S. Sakai, Phys. Rev. Lett. 94 (2005) 072305;
Nucl. Phys. A 774 (2006) 775.

\bibitem{Cass99} E.~L.~Bratkovskaya and W.~Cassing, Nucl. Phys. A
619 (1997) 413;   W.~Cassing and E.~L.~Bratkovskaya, Phys.~Rept.~{
308} (1999)  65.

\bibitem{URQMD1}
    S.A.~Bass {\it et al.},
    Prog. Part. Nucl. Phys. {42} (1998) 279.

\bibitem{URQMD2}
    M.~Bleicher {\it et al.},
    J. Phys. G {25} (1999) 1859.

\bibitem{Geiger} K. Geiger, Phys. Rep. 258 (1995) 237.

\bibitem{Zhang} B. Zhang, M. Gyulassy, and C. M. Ko, Phys. Lett. B
455 (1999) 45.

\bibitem{Molnar} D. Molnar and M. Gyulassy, Phys. Rev. C 62 (2000) 054907;
Nucl. Phys. A 697 (2002) 495; Nucl. Phys. A 698 (2002) 379.

\bibitem{Bass} S. A. Bass, B. M\"uller, and D. K. Srivastava, Phys. Lett. B 551 (2003) 277;
Acta Phys. Hung. A 24 (2005) 45.

\bibitem{AMPT} Z.-W. Lin {\it et al.}, Phys. Rev. C 72 (2005)
064901.

\bibitem{Carsten} Z. Xu and C. Greiner, Phys. Rev. C 71 (2005) 064901;
Nucl. Phys. A 774 (2006) 034909.

\bibitem{Juchem} W. Cassing and S. Juchem, Nucl. Phys. A 665
(2000) 417; Nucl. Phys. A 672 (2000) 417.

\bibitem{Sascha1} S. Juchem, W. Cassing, and C. Greiner, Phys. Rev. D 69 (2004) 025006;
 Nucl. Phys. A 743 (2004) 92.

\bibitem{Leo} S. Leupold, Nucl. Phys. A 672 (2000) 475.

\bibitem{Laura} W. Cassing, L. Tolos, E. L. Bratkovskaya, and A. Ramos,
Nucl. Phys. A 727 (2003) 59.

\bibitem{Andre04} A. Peshier, Phys. Rev. D 70 (2004) 034016.

\bibitem{Andre05} A. Peshier, J. Phys. G 31 (2005) S371.

\bibitem{Karsch5} F. Karsch, Nucl. Phys. A 698 (2002) 199c; F.
Karsch, E. Laermann and A. Peikert, Phys. Lett. B 478 (2000) 447.

\bibitem{BlaizIR}
  J.\,P.~Blaizot, E.~Iancu, and A.~Rebhan, Phys.\ Rev.\ D 63 (2001) 065003.

\bibitem{pQP}
  A.~Peshier, B.~K{\"a}mpfer, O.\,P.~Pavlenko, and G.~Soff,
  Phys.\ Rev.\ D 54 (1996) 2399;
  P.~Levai, U.~Heinz, Phys.\ Rev.\ C 57 (1998) 1879;
  A.~Peshier, B.~K{\"a}mpfer, G.~Soff, Phys.~Rev.~C 61 (2000) 045203,
  Phys.~Rev.~D 66 (2002) 094003.

\bibitem{Rafelski} J. Letessier and J. Rafelski, Phys. Rev. C 67 (2003)
031902.

\bibitem{Bielefeld} O. Kaczmarek, F. Karsch, F. Zantow, and P.
Petreczky, Phys. Rev. D 70 (2004) 074505; erratum-ibid. D 72
(2005) 059903.

\bibitem{Pisar89LebedS}
 R.\,D.~Pisarski, Phys.\ Rev.\ Lett.\ 63 (1989) 1129;
 V.\,V.~Lebedev and A.\,V.~Smilga, Ann.\ Phys.\ (N.Y.) 202 (1990) 229.

\bibitem{Peshi}
  A.~Peshier, Phys.\ Rev.\ D63 (2001) 105004.

\bibitem{CCPACS}
 M.~Okamoto {\em et al.}, Phys.\ Rev.\ D 60  (1999) 094510.

\bibitem{Dosch} G. Boyd {\em et al.}, Nucl. Phys. B 469 (1996)
419.

\bibitem{excita} W. Cassing, E. L. Bratkovskaya, and S. Juchem,
Nucl. Phys. A 674 (2000) 249.


\bibitem{Cass07} W. Cassing, to be published

\bibitem{Toneev} Yu.B.~Ivanov, V.V.~Skolov, and V.D.~Toneev, Phys.
Rev. D 71 (2005) 014005.

\bibitem{SIGMAM} B. D. Serot and J. D. Walecka, Adv. Nucl. Phys.
16 (1986) 1; B. D. Serot, Rep. Prog. Phys. 55 (1992) 1855; P. G.
Reinhard, Rep. Prog. Phys. 52 (1989) 439.

\bibitem{PHSD} W. Cassing, talk at ECT$^*$, Workshop on {\it
Parton Propagation through Strongly Interacting Matter}, September
27, 2005, [http://conferences.jlab.org/ECT/program].

\bibitem{Lenske} C. Fuchs, H. Lenske, and H. H. Wolter, Phys. Rev.
C 52 (1995) 3043.


\bibitem{Ratti} C. Ratti, M. A. Thaler and W. Weise, Phys. Rev. D
73 (2006) 014019; C. Ratti and W. Weise, Phys. Rev. D 70 (2004)
054013.




\end{thebibliography}
\end{document}